\documentclass{article}
\usepackage{amssymb}
\usepackage{amsfonts}
\usepackage{amsmath}
\usepackage{latexsym}

\begin{document}

\title{Density perturbations in the gas of wormholes}
\author{A.A. Kirillov, E.P. Savelova \\
%EndAName
\emph{Uljanovsk State University, Branch in Dimitrovgrad, }\\
\emph{Dimitrova str 4.,} \emph{Dimitrovgrad, 433507, Russia} }
\date{}
\maketitle

\begin{abstract}
The observed dark matter phenomenon is attributed to the presence of a gas
of wormholes. We show that due to topological polarization effects the
background density of baryons generates non-vanishing values for wormhole
rest masses. We infer basic formulas for the scattering section between
baryons and wormholes and equations of motion. Such equations are then used
for the kinetic and hydrodynamic description of the gas of wormholes. In the
Newtonian approximation we consider the behavior of density perturbations
and show that at very large distances wormholes behave exactly like heavy
non-baryon particles, thus reproducing all features of CDM models. At
smaller scales (at galaxies) wormholes strongly interact with baryons and
cure the problem of cusps. We also show that collisions of wormholes and
baryons lead to some additional damping of the Jeans instability in baryons.
\end{abstract}

\section{Introduction}

The nature of Dark Matter (DM) represents one of the most important and yet
unsolved problems of the modern astrophysics. Indeed, while the presence of
DM has long been known \cite{Zw} and represents a well established fact
(e.g., see \cite{dm1,Pr} and references therein), there is no common
agreement about what DM is. In the simplest picture DM represents some
non-baryonic particles (predicted numerously by particle physics) which
should be sufficiently heavy to be cold at the moment of recombination and
those give the basis to the standard (cold dark matter) CDM models. The
latter turn out to be very successful in reproducing properties of the
Universe at very large scales (where perturbations are still on the linear
stage of the development) which led to a wide-spread optimistic believe that
non-baryonic particles provide indeed an adequate content of DM.

However the success of CDM models at very large scales is accompanied with a
failure at smaller (of the galaxies size) scales. Indeed, cold particles
which interact only by gravity should necessary form cusps ($\rho _{DM}\sim
1/r$) in centers of galaxies\footnote{%
The presence of cusps formed by the development of adiabatic perturbations
follows straightforwardly from the conservation of the circulation theorem
in the hydrodynamics. By other words the fact that the distribution of DM
should have cusps in galaxies is equivalent to the fact that DM should
represent cold non-baryonic particles.} \cite{NFW} (see also \cite{Cusp}
where the problem of cusps in CDM is discussed in more detail), while
observations \cite{Core} definitely show the cored ($\rho _{DM}\sim const$)
distribution. The only way to destroy the cusp and get the cored
distribution is to introduce some self-interaction in DM or to consider warm
DM. Both possibilities are rejected at large scales by observing $\Delta T/T$
spectrum (e.g., see \cite{Pr} and references therein). By other words DM
displays so non-trivial properties (it is warm or self-interacting in
galaxies, however it was cold at the moment of recombination and it is still
cold on larger (than galaxies) scales) that it is difficult to find
particles capable of reconciling such observations.

These facts support the constant interest to different alternatives of the
DM hypothesis which interpret the observed discrepancy between luminous and
gravitational masses as a violation of the law of gravity. Such violations
(or modifications of general relativity (GR)) have widely been discussed,
e.g., see \cite{VN,MOND}. However, it turns out to be rather difficult to
get a modification of GR which is flexible enough to reconcile all the
variety of the observed DM halos. Moreover, the weak lensing observations of
a cluster merge \cite{DMpaper} seem to reject most of modifications of GR in
which a non-standard gravity force scales with baryonic mass.

The more viable picture of DM phenomena was suggested by \cite{K06} (see
also references therein) and developed recently by \cite{KT06} \cite{KSS09}.
It is based on the fact that on the very early (quantum) stage the Universe
should have a foam-like topological structure \cite{wheeler}. There are no
convincing theoretical arguments of why such a foamed structure should decay
upon the quantum stage - relics of the quantum stage foam might very well
survive the cosmological expansion, thus creating a certain distribution of
wormholes in the Friedman space. Moreover, the inflationary stage in the
past \cite{inf} should enormously stretch characteristic scales of the relic
foam. The foam-like structure, in turn, was shown to be flexible enough to
account for the all the variety of DM phenomena \cite{K06,KT07}; for
parameters of the foam may arbitrary vary in space to produce the observed
variety of DM halos in galaxies (e.g., the universal rotation curve for
spirals constructed by \cite{KT06} for the foamed Universe perfectly fits
observations). Moreover, the topological origin of DM phenomena means that
the DM halos surrounding point-like sources appear due to the scattering on
topological defects and if a source radiates, such a halo turns out to be
luminous too \cite{KSS09} which seems to be the only way to explain
naturally the observed absence of DM fraction in intracluster gas clouds
\cite{DMpaper}.

The foam-like structure of the Universe is represented by the gas of
wormholes randomly distributed in space. It was demonstrated recently by
\cite{KS07} that in the presence of such a gas every point source turns out
to be surrounded with a "dark halo" which possesses both signs depending on
scales and the background distribution of wormholes. Therefore, it seems to
be not quite clear for readers whether wormholes produce the necessary CDM
picture, or they merely suggest some additional effects. Moreover, there
still exists some wide-spread mistaken opinion that wormholes lead to
non-Gaussian perturbations in the analogy with topological defects of
another kind (strings, monopoles, etc.). In the present paper we clarify
such problems by means of considering the development of density
perturbations in the gas of wormholes. We demonstrate that at very large
scales wormholes behave exactly like very heavy particles and thus
reproducing all the predictions of CDM. They however predict an additional
specific damping in the development of baryon perturbations. Moreover, at
smaller scales the non-linear stage of the evolution of perturbations
essentially diverges from that in CDM. At small scales there exists a rather
strong non-gravitational interaction between particles and wormholes (due to
the mutual scattering) which surely cure the problem of cusps in galaxies.

The complete analysis of the gravitational dynamics of wormholes is rather
complicated. Therefore, in the present paper we restrict ourself with the
Newton approximation to derive basic equations which govern the dynamics of
baryons and wormholes. While the problem of the generalization to the
General Relativity we leave for the future research.

The paper is organized as follows. In Sec.2 we introduce a wormhole and
describe it's general properties. In Sec. 3 we show that in the Friedman
model wormholes acquire non-vanishing rest masses. In Sec. 4 we describe the
process of the scattering of particles on a wormhole. In Sec. 5 we infer the
motion equations for particles and wormholes in the Newtonian approximation
for the expanding reference system. In Sec.6 we introduce the system of
Boltzmann - Vlasov equations which describes kinetics of particles,
wormholes, and collisions. In Sec. 7 we infer the non-relativistic
hydrodynamic equations with corrections for the collisions between particles
and wormholes. In Sec.8 we calculate basic kinetic coefficients which
describe the collisions. In Sec.9 we consider the behavior of linear
perturbations in the Newtonian approximation. In particular, we explicitly
demonstrate that at large distances wormholes behaves exactly like heavy
non-baryon particles reproducing thus the standard CDM picture. While the
strong coupling at smaller scales cures the basic failure of CDM (i.e., it
removes cusps in galaxies). In the last section we discuss results obtained
and show further perspectives.

\section{Wormholes}

The simplest wormhole is described by the metric%
\begin{equation}
ds^{2}=c^{2}dt^{2}-h^{2}\left( r\right) \delta _{\alpha \beta }dx^{\alpha
}dx^{\beta },  \label{wmetr}
\end{equation}%
where
\begin{equation}
h\left( r\right) =1+\theta \left( b-r\right) \left( \frac{b^{2}}{r^{2}}%
-1\right)
\end{equation}%
and $\theta \left( x\right) $ is the step function ($\theta \left( x\right)
=0$ as $x<0$ and $\theta \left( x\right) =1$ as $x>0$ ). We point out the
obvious properties $h_{r}^{\prime }=\theta \left( b-r\right) \left( \frac{%
b^{2}}{r^{2}}\right) ^{\prime }$ and $h_{rr}^{\prime \prime }=\delta \left(
b-r\right) \left( \frac{b^{2}}{r^{2}}\right) ^{\prime }\left( b-r\right)
^{\prime }+\theta \left( b-r\right) \left( \frac{b^{2}}{r^{2}}\right)
^{\prime \prime }$. Such a wormhole has vanishing throat length. Indeed, for
the region $r>b$ $h=1$ and the metric coincides merely with that in the
Minkowsky space, while the region $r<b$, upon the obvious transformations $%
y^{\alpha }=\frac{b^{2}}{r^{2}}x^{\alpha }$, gives the same region $y>b$
with the same flat metric $ds^{2}=c^{2}dt^{2}-\delta _{\alpha \beta
}dy^{\alpha }dy^{\beta }$. Therefore, both regions $r>b$ and $r<b$ represent
similar portions of two Minkowsky spaces glued at the surface of the sphere $%
S^{3}$ with the center at the origin $r=0$ and the radius $r=b$. Thus,
formally, such a space can be described with the ordinary double-valued flat
metric in the region $r_{\pm }>b$ as%
\begin{equation}
ds^{2}=c^{2}dt^{2}-\delta _{\alpha \beta }dx_{\pm }^{\alpha }dx_{\pm
}^{\beta },  \label{wmetr2}
\end{equation}%
where the sign $\pm $ in coordinates $x_{\pm }^{\alpha }$ stands to describe
two different sheets of space.

A generalization appears when we change the step function $\theta \left(
x\right) $ with any smooth function $\widetilde{\theta }\left( x\right) $
which has the same property $\widetilde{\theta }\left( x\right) \rightarrow
0 $ as $x\rightarrow -\infty $ and $\widetilde{\theta }\left( x\right)
\rightarrow 1$ as $x\rightarrow 0$. On the contrary to the previous case
such a wormhole will have a non-vanishing throat length. However in the last
case the consideration differs in details, while general features remain the
same. We also point out that such wormholes have vanishing mass. In general,
one may also insert a non-vanishing mass to the wormhole as it is described
in \cite{Vis}.

First of all we consider the stress energy tensor which produces such a
wormhole. It can be found from the Einstein equation $T_{\alpha }^{\beta
}=R_{\alpha }^{\beta }-\frac{1}{2}\delta _{\alpha }^{\beta }R$. Since the
metric (\ref{wmetr}) does not depend on time we find
\begin{equation}
R_{0}^{0}=R_{\alpha }^{0}=0,\ \ \ R_{\alpha }^{\beta }=-P_{\alpha }^{\beta
}=-\frac{2}{b}\delta \left( b-r\right) \{n_{\alpha }n^{\beta }-\delta
_{\alpha }^{\beta }\}
\end{equation}%
where $n^{\alpha }=n_{\alpha }=x^{\alpha }/r$ is the outer normal to the
sphere $S^{3}$ and therefore
\begin{equation}
T_{\mu }^{\nu }=-\frac{2}{b}\delta \left( b-r\right) \left( \delta _{\mu
}^{\nu }-\Delta _{\mu }^{\nu }\right)
\end{equation}%
where $\Delta _{0}^{0}=\Delta _{\alpha }^{0}=0$, and $\Delta _{\alpha
}^{\beta }=\delta _{\alpha }^{\beta }-n_{\alpha }n^{\beta
}=\sum_{A=1,2}e_{\alpha }^{A}e^{A\beta }$ where $e_{\alpha }^{A}$ are two
unite tangent to the sphere vectors $e_{\alpha }^{A}n^{\alpha }=0$. Thus,
the effective source can be considered as a mixture of a negative
"cosmological constant" $T_{\mu }^{0\nu }=-\frac{2}{b}\delta \left(
b-r\right) \delta _{\mu }^{\nu }$ and two components of a perfect fluid $%
T_{\mu }^{A\nu }=-\frac{2}{b}\delta \left( b-r\right) u_{\mu }^{A}u^{A\nu }$
with zero pressure $p=0$, energy density $\varepsilon =-\frac{2}{b}\delta
\left( b-r\right) $, and the velocity $u^{A\nu }=\left( 0,e^{A\beta }\right)
$. The velocities have purely space-like (tangent to the surface of the
sphere) components $u_{\mu }^{A}u^{A\mu }=-1$. This matter is concentrated
only on the surface of the sphere due to the multiplier $\delta \left(
b-r\right) $. All such sources represent an exotic matter which cannot be
constructed from actual particles. Recall, that all real particles have
time-like velocities $u_{\mu }u^{\mu }=1$ or isotropic $u_{\mu }u^{\mu }=0$
(if the rest mass vanishes $m=0$). However, this property is violated for
virtual particles. This gives us a hint that vacuum polarization effects can
in principle be collected to organize such a form of matter. By other words
to construct a wormhole we have to organize a negative cosmological term on
the surface $S^{3}$ and then to rotate the surface with the space-like
velocity in the two different directions. From the other hand the presence
of a wormhole leads to the vacuum polarization and therefore to the exotic
matter (e.g., see \cite{Grib,whst}). Here we leave aside the important
problem of what precedes the egg (wormhole) or the hen (exotic matter).

The inner and outer regions of the sphere $S^{3}$ are equal, which gives the
possibility to construct a wormhole which connects regions in the same space
(instead of two independent spaces). This is achieved by the the
identification (gluing) in (\ref{wmetr2}) of the two spaces by means of the
use of possible motions of the flat space. Let $\vec{R}_{+}$ be the position
of the sphere in coordinates $x_{+}^{\alpha }$, then the gluing is the rule%
\begin{equation}
x_{+}^{\alpha }=R_{+}^{\alpha }+U_{\beta }^{\alpha }\left( x_{-}^{\beta
}-R_{-}^{\beta }\right)  \label{gl}
\end{equation}%
where $U_{\beta }^{\alpha }\in O(3)$, which represents the composition of a
translation and a rotation of the flat space. In terms of common coordinates
such a wormhole represents the standard flat space in which the two spheres $%
S_{\pm }^{3}$ (with centers at positions $R_{\pm }^{\alpha }$) are glued by
the rule (\ref{gl})

If we neglect the higher order images with respect to the transformations $%
\xi _{\pm }^{\alpha }=\frac{b^{2}}{\xi _{\pm }^{2}}\xi _{\pm }^{\alpha }$
(which are the reflections with respect to spheres $S_{\pm }^{3}$), then we
can use the metric (\ref{wmetr}) in which we merely replace $h$ with the
function
\begin{equation}
h\left( r,R_{\pm }\right) =1+\theta \left( b-\left\vert \xi _{+}\right\vert
\right) \left( \frac{b^{2}}{\xi _{-}^{\prime 2}}-1\right) +\theta \left(
b-\left\vert \xi _{-}\right\vert \right) \left( \frac{b^{2}}{\xi
_{+}^{\prime 2}}-1\right)
\end{equation}%
where $\xi _{\pm }^{\alpha }=x^{\alpha }-R_{\pm }^{\alpha }$ and $\xi _{\pm
}^{\prime \alpha }=(U_{\beta }^{\alpha })^{\pm 1}\xi _{\pm }^{\beta }$. In
this case the physically admissible region of space is the outer region of
the two spheres $S_{\pm }^{3}$, while the inner regions represent only
additional images of the outer space and have no meaning\footnote{%
If we do not neglect the higher order images, then within every sphere will
appear a countable set of spheres $S_{A\pm }^{3}$ which represent additional
images (copies) of the outer (physically admissible) region of space.}.

Thus, we see that the simplest wormhole may be described by a set of
parameters $\eta =\left( R_{+}^{\alpha },R_{-}^{\alpha },b,U_{\beta
}^{\alpha }\right) $. We point out that in the general case a wormhole
possesses a rather complex properties, e.g., parameters possess a dynamics $%
\eta =\eta \left( t\right) $ and, moreover, ($b,U_{\beta }^{\alpha }$) are
functions of coordinates $x^{\alpha }$ which describe the specific structure
of the wormhole.

The generalization to a set of wormholes is given by
\begin{equation}
h\left( r,R_{\pm }\right) =1+\sum\limits_{n,\sigma =\pm }\theta \left(
b_{n}-\left\vert \xi _{n\sigma }\right\vert \right) \left( \frac{b_{n}^{2}}{%
\xi _{n-\sigma }^{2}}-1\right)  \label{h}
\end{equation}%
and the identification of points at joint spheres $S_{n\pm }^{3}$ as $\xi
_{n+}^{\alpha }=U_{n\beta }^{\alpha }\xi _{n-}^{\beta }$.

\section{The rest mass of a wormhole in the expanding Universe}

As it was demonstrated previously (e.g., see for details \cite{KS07}) the
presence of a gas of wormholes leads to a specific topological
polarizability of space. This can be described as a bias of point-like
sources of gravity%
\begin{equation*}
\delta (r-r_{0})\rightarrow \delta (r-r_{0})~+B\left( r,r_{0}\right) ,
\end{equation*}%
where the bias gives the polarization mass generated on throats. The bias
consists of two terms $B=B_{0}+B_{1}$, where $B_{0}$ resembles the bias of
spherical mirrors and gives the positive contribution (anti-screening) to
the total mass of a particle
\begin{equation}
B_{0}\left( r\right) =\sum_{n,\sigma =\pm }\frac{b_{n}}{R_{\sigma }}\left[
\delta (\vec{r}-\vec{r}_{n,\sigma })-\delta (\vec{r}-\vec{R}_{n,\sigma })%
\right] ,  \label{bk0}
\end{equation}%
where $\vec{r}_{n,\pm }=\vec{R}_{n,\pm }+\frac{a^{2}}{(\vec{r}_{0}-\vec{R}%
_{\mp })^{2}}U_{n}^{\pm 1}(\vec{r}_{0}-\vec{R}_{n,\mp })$, while the rest
part gives pure screening and is given by
\begin{equation}
B_{1}\left( r\right) =\sum_{n,}b_{n}\left( \frac{1}{R_{n,+}}-\frac{1}{R_{n,-}%
}\right) \left[ \delta (\vec{r}-\vec{r}_{n,-})-\delta (\vec{r}-\vec{r}_{n,+})%
\right] .  \label{bk1}
\end{equation}%
In the homogeneous Universe (i.e., in the case of a homogeneous distribution
of matter) the second part disappears automatically, while the first part
generates the rest mass for every wormhole. Indeed, to demonstrate this we
rewrite the bias (\ref{bk0}) in the form (see for details \cite{KS07})
\begin{equation}
B_{0}\left( r\right) =\frac{\partial h\left( \vec{r}\right) }{\partial
r^{\alpha }}\frac{\partial \left( -1/r\right) }{\partial r^{\alpha }}+4\pi
h\left( 0\right) \delta \left( \vec{r}\right) ,  \label{bm}
\end{equation}%
where
\begin{equation*}
h\left( r\right) =\int \frac{r^{\alpha }R^{\beta }}{R^{2}}\left[ H_{\alpha
\beta }^{+}\left( \vec{r},\vec{R}\right) +H_{\beta \alpha }^{+}\left( \vec{R}%
,\vec{r}\right) \right] d^{3}R,
\end{equation*}%
\begin{equation*}
H_{\alpha \beta }^{\pm }\left( R_{+,}R_{-}\right) =\int b^{3}U_{\alpha \beta
}^{\pm 1}F\left( R_{\pm },b,U\right) dadU,
\end{equation*}%
and $F\left( \eta \right) $ is the distribution of wormholes in the
configuration space. In the case of the homogeneous and isotropic
distribution of wormholes, averaging over the rotation matrix gives $%
H_{\alpha \beta }^{\pm }\left( R_{+,}R_{-}\right) =\frac{1}{3}\delta
_{\alpha \beta }\phi \left( \left\vert R_{+}-R_{-}\right\vert \right) $,
where $\phi \left( X\right) =\int b^{3}F\left( X,b\right) db$. Therefore,
the bias reduces to the form
\begin{equation}
B_{0}\left( r,r_{0}\right) =\frac{4\pi }{3}n\overline{b^{3}}\delta \left(
\vec{r}-\vec{r}_{0}\right)
\end{equation}%
where $n$ is the density of throats, while $\frac{4\pi }{3}n\overline{b^{3}}$
is the portion of the unit volume which is cut by throats. In the above
formula the only difference from spherical mirrors (the multiplier $1/3$)
appears due to averaging over the rotation matrix. By other words every
wormhole increases the mass of a point-like source in proportion to the
volume which the wormhole cuts from the space. When the space is filled with
homogeneously distributed matter, every wormhole acquires the rest mass $%
M_{w}\left( b\right) =\frac{4\pi }{3}b^{3}\rho $ ($\rho $ is the mean
density).

All the above consideration can be omitted if we note that the wormhole
possesses straightforward generalization to the case of the expanding
Universe. Indeed, the metric is%
\begin{equation}
ds^{2}= a^{2}\left( \tau \right) \left( d\tau ^{2}-h^{2}\left( r\right)
\delta _{\alpha \beta }dx^{\alpha }dx^{\beta }\right) .  \label{fr}
\end{equation}%
In this case the wormhole expands with the space and acquires a
non-vanishing rest mass
\begin{equation}
M_{w}=\frac{4}{3}\pi a^{3}b^{3}\rho _{b}  \label{M}
\end{equation}%
where $\rho _{b}$ is the mean density of matter in the Universe. By other
words the gluing procedure cuts two equal homogeneous portions of space (the
two spheres $S_{\pm }^{3}$) filled with a homogeneously distributed matter.
Thus, the total amount of matter diminishes, while spheres acquire the rest
masses which compensate the distribution of the effective density of matter
to the homogeneity\footnote{%
We also point out that rotating wormholes will be described by the tensor of
inertia $I_{\alpha \beta }^{w}=I\delta _{\alpha \beta }$, with $I=\frac{2}{5}%
a^{2}b^{2}M_{w}$.}. We point out that the metric (\ref{fr}), with (\ref{h})
taken into account, directly shows that an arbitrary primordial distribution
of wormholes in space agrees with the visible homogeneity of space (the
Friedman model) as it was first stated by \cite{K06}. An inhomogeneous
distribution of wormholes in space (e.g., fractal distribution) leads to the
dual fractal distribution of particles (for particles may occupy only the
physically admissible region of space), while the total effective density
remains perfectly homogeneous.

From the astrophysical standpoint such a mass is much smaller than that of a
typical object (e.g., if the throat radius is $r_{th}=ab\sim R_{0}$, and $%
\rho _{b}=$ $\rho _{cr}$, where $R_{0}$ is the Sun radius and $\rho _{cr}$
is the critical density, then the mass has the order $M_{w}\sim 10kg$).
Thus, gravitational effects of such objects in Solar systems is merely
negligible. However in the dynamics of larger systems (galaxies, clusters,
etc.) they become the more and more visible , e.g., as the presence of dark
matter. We point out that such dark particles (wormholes) are extremely
heavy from the particle physics point of view. It is clear that the picture
where the Universe is filled with a gas of "particles" whose rest masses
have the order $\sim 10kg$ perfectly fits the basic Cold Dark Matter model
(CDM) widely accepted today. And as it is well known, predictions of CDM
perfectly agree with observations at very large (clusters of galaxies and
higher) scales. However, while the standard CDM (heavy non-baryon particles)
makes a wrong prediction at smaller scales (e.g., it predicts cusps $\rho
_{DM}\sim 1/r$ in centers of galaxies, when observations demonstrate the
cored $\rho _{DM}\sim const$ distribution), we may expect that wormholes
will cure such a problem. Indeed, since the two conjugated spheres $S_{\pm
}^{3}$ (throats) represent merely the "same" region of space we may state
that the local density at $S_{+}^{3}$ coincides exactly with that at $%
S_{-}^{3}$. Let $d_{0}$ be the typical distance between throats. Thus, if
one throat gets into the central region of a galaxy while the rest throat is
sufficiently far from the center, then the total density will be somewhat
smoothed. And the typical scale of the smoothing (i.e. the minimal core
radius) will be of the order of $d_{0}$. Of course, the rigorous
consideration of this problem requires considering the proper dynamics of
wormholes which we start in the next section.

\section{The collision with a wormhole}

Consider the simplest construction of a wormhole in space as follows. Let us
fold the space at the plane $z=0$. Then we can use the simplest metric (\ref%
{wmetr}) which connects the two half-spaces $z>0$ and $z<0$. Let the
velocity of the wormhole is $\vec{V}$ \ which, due to the above symmetry
(i.e., $Z_{+}=-Z_{-}$, where $\vec{R}_{\pm }=(X_{\pm },Y_{\pm },Z_{\pm })$
are the positions of throats), corresponds to the case when one throat (say $%
S_{-}^{3}$) moves with the velocity $\vec{V}_{-}=\vec{V}$, while the other $%
S_{+}^{3}$ moves with $\vec{V}_{+}=(V_{x},V_{y},-V_{z})$. For the sake of
simplicity we consider everywhere the non-relativistic case (i.e., $V\ll c$%
). In the case of a general wormhole the relation between $\vec{V}_{+}$ and $%
\vec{V}_{-}$ is given by the same relation (\ref{gl})%
\begin{equation}
V_{+}^{\alpha }=U_{\beta }^{\alpha }V_{-}^{\beta }.  \label{vr}
\end{equation}%
By other words since both spheres represent the same region of space their
velocities are rigidly connected (the visible doubling of the number of
degrees of freedom related to wormholes is fictitious).

Consider an incident on $S_{-}^{3}$ particle with the rest mass $m$ and the
initial velocity $\vec{v}$. Then the scattering of the particle leads to the
transformation\footnote{%
This law of the transformation can be easily obtained as follows. We notice
that the metric (\ref{wmetr}) merely connects two equal flat spaces. Then
the scattering of a particle merely coinsides with the elastic scattering on
a solid ball with simultaneous transport of the particle from one sheet of
space to the other sheet.}
\begin{equation}
n_{-}^{\alpha }\rightarrow n_{+}^{\prime \alpha }\ =U_{\beta }^{\alpha
}n_{-}^{\beta },\ \ \ \vec{R}_{\pm }\rightarrow \vec{R}_{\pm }^{\prime }=%
\vec{R}_{\pm }  \label{tr1}
\end{equation}%
where $n_{\pm }^{\alpha }=\left( x^{\alpha }-R_{\pm }^{\alpha }\right) /b$
are points at spheres $S_{\pm }^{3}$, and the transformation of velocities%
\begin{equation}
\vec{V}_{\pm }^{\prime }=\vec{V}_{\pm }+\frac{2m}{M_{w}+m}\left( u_{\pm
}n_{\pm }\right) \vec{n}_{\pm },\   \label{tr2}
\end{equation}%
\begin{equation}
v^{\prime \alpha }=U_{\beta }^{\alpha }\left( v^{\beta }-\frac{2M_{w}}{%
M_{w}+m}\left( u_{-}n_{-}\right) n_{-}^{\beta }\right) =  \label{tr3}
\end{equation}%
\begin{equation*}
=V_{+}^{\alpha }+u_{+}^{\alpha }-\frac{2M_{w}}{M_{w}+m}\left(
u_{+}n_{+}\right) n_{+}^{\alpha }.
\end{equation*}%
Here $(un)=(\vec{u}\vec{n})$ denotes the ordinary scalar product and
\begin{equation*}
\vec{u}_{-}=\vec{v}-\vec{V}_{-},\ \ \ u_{+}^{\alpha }=U_{\beta }^{\alpha
}u_{-}^{\beta },
\end{equation*}%
\newline
and the rotational matrix $U_{\beta }^{\alpha }\in O(3)$ corresponds to the
above symmetry (i.e., the reflection with respect to the plane $z=0$, $%
z^{\prime }=-z$).

We point out that the above law of transformation (\ref{tr1})-(\ref{tr3})
holds in the most general case (the case of an arbitrary gluing, i.e., an
arbitrary $U_{\beta }^{\alpha }\in O(3)$) as well. In the limit $%
M_{w}\rightarrow \infty $ this law has been considered by us in \cite{KSZ}.

It is easy to see that the law of transformation (\ref{tr1})-(\ref{tr3})
conserves the total energy%
\begin{equation*}
mv^{2}+MV_{-}^{2}=mv^{\prime 2}+MV_{+}^{\prime 2}
\end{equation*}%
where the kinetic energy of the wormhole is accounted only for once. It
maybe convenient to consider throats as an independent objects. Then, due to
the rigid relation (\ref{vr}) we have to introduce the bound energy $%
\varepsilon =-\frac{1}{2}MV_{-}^{2}=-\frac{1}{2}MV_{+}^{2}$ which gives the
conservation of the total energy in the form%
\begin{equation}
\frac{mv^{2}}{2}+\frac{MV_{-}^{2}}{2}+\frac{MV_{+}^{2}}{2}+\varepsilon =%
\frac{mv^{\prime 2}}{2}+\frac{MV_{-}^{\prime 2}}{2}+\frac{MV_{+}^{\prime 2}}{%
2}+\varepsilon ^{\prime }\text{.}  \label{ce}
\end{equation}

It is important that if we forget about the factorization (i.e., the
identification of points or the gluing) (\ref{gl}), then the above
transformation does not conserve the momentum. Instead we get the
conservation in the form
\begin{equation}
MV_{+}^{\prime \alpha }+mv^{\prime \alpha }=U_{\beta }^{\alpha }\left(
MV_{-}^{\alpha }+mv^{\alpha }\right) .  \label{cm}
\end{equation}%
It happens due to the fact that the space with the wormhole does not possess
the translation symmetry. It possesses however the somewhat entangled (with
the gluing (\ref{gl}) taken into account) symmetry.

We point out that when we consider dynamics of wormholes the matrix $%
U_{\beta }^{\alpha }$ becomes function on time (formation of gravitationally
bounded objects leads naturally to origin of rotational motions).

\section{Equations of motion}

As it was shown above a general wormhole possesses a rest mass and,
therefore, it should move in space as an ordinary test particle. For the
sake of simplicity we shall use an approximation when wormholes can be
considered as point-like objects, i.e., we neglect the size of wormholes.
For cosmological applications this represents a rather good approximation,
while for some astrophysical problems it can be not sufficient.

The presence of a gas wormholes in the Universe (i.e. of the complexity of
the topological structure of space) leads to an enormous complexity in
consideration of the dynamics (due to boundary conditions at wormholes). To
simplify the problem we shall use the Newton's equations, while the
generalization to the relativistic case we leave for the future. Moreover,
as it is well known (e.g., see  \cite{Pbl}) the Newton's approach is
acceptable for a rather huge range of scales. It will be also convenient to
introduce the expanding reference system from the very beginning.

\subsection{Newton's equation}

When the gravitational field is rather weak, the Einstein equations reduce to

\begin{equation}
\nabla _{r}^{2}\Phi =4\pi G\left( \rho +\frac{3P}{c^{2}}\right)  \label{1}
\end{equation}%
where $\nabla _{r}^{2}\Phi \approx R_{00}$ is the time component of the
Ricci tensor. The equations of motion for a test particle are
\begin{equation}
\frac{d^{2}r^{\alpha }}{dt^{2}}=-\Phi _{,\alpha }  \label{2}
\end{equation}%
which have the range of the applicability $v<<c$, or (equivalently) at
distances $R<<cH^{-1}$($H=\dot{a}/a$ is the Hubble constant). The lower
boundary is given by the gravitational radius for compact objects (black
holes).

Consider the homogeneous component $P(t)$ and $\rho (t)$ and the homogeneous
expansion of space $\vec{r}=a(t)\vec{x}$. Then the equation (\ref{1}) gives
\begin{equation*}
\Phi _{b}=\frac{2}{3}\pi G\left( \rho _{b}(t)+\frac{3P_{b}(t)}{c^{2}}\right)
r^{2},
\end{equation*}%
while the equations of motion (\ref{2}) lead to the standard cosmological
equation
\begin{equation}
\frac{d^{2}a}{dt^{2}}=-\frac{4}{3}\pi G\left( \rho _{b}(t)+\frac{3P_{b}(t)}{%
c^{2}}\right) a.  \label{coseq}
\end{equation}%
Here the point $x=0$ corresponds to the observer (which gives $\Phi _{b}\sim
r^{2})$ and the applicability is restricted by
\begin{equation*}
g_{00}=1+2\Phi /c^{2}\qquad 2\Phi <<c^{2}
\end{equation*}%
which means that distances $r$ cannot be too large.

\subsection{Peculiar velocity}

Consider now the expanding reference system $\vec{r}=a(t)\vec{x}$. Then the
total velocity is
\begin{equation*}
\vec{u}=a\dot{\vec{x}}+\vec{x}\dot{a}=a\dot{\vec{x}}+H\vec{r}
\end{equation*}%
where $a\dot{\vec{x}}$ is the peculiar velocity, while $H\vec{r}$ gives the
standard Hubble expansion.

For a particle the equations of motion can be obtained from the Lagrangian (%
\cite{Pbl})
\begin{equation*}
\mathcal{L}=\frac{1}{2}m\left( a\dot{\vec{x}}+\vec{x}\dot{a}\right)
^{2}-m\Phi (\vec{x},t).
\end{equation*}%
Let us use the canonical transformation
\begin{equation*}
\mathcal{L}\rightarrow \mathcal{L}-\frac{d\psi }{dt},\psi =\frac{1}{2}m\dot{a%
}ax^{2}
\end{equation*}%
which transforms the Lagrangian function to the form%
\begin{equation*}
\mathcal{L}=\frac{1}{2}ma^{2}\dot{\vec{x}}^{2}-m\varphi ,
\end{equation*}%
where
\begin{equation*}
\varphi =\Phi +\frac{1}{2}a\ddot{a}x^{2}
\end{equation*}%
and new Newtonian potential $\varphi $ satisfy the equation%
\begin{equation*}
\frac{1}{a^{2}}\nabla ^{2}\varphi =4\pi G\left( \rho +\frac{3P}{c^{2}}%
\right) +3\frac{\ddot{a}}{a}
\end{equation*}%
where the gradient is taken over $\vec{x}$. Then by means of use of the
cosmological equation (\ref{coseq}) it transforms the Poisson equation to
the form%
\begin{equation*}
\frac{1}{a^{2}}\nabla ^{2}\varphi =4\pi G\left( \delta \rho (\vec{x},t)+%
\frac{3\delta P(\vec{x},t)}{c^{2}}\right) ,
\end{equation*}%
where $\delta \rho =\rho (\vec{x},t)-\rho _{b}(t)$. The equations of motions
for a test particle become
\begin{equation}
\vec{p}=ma^{2}\dot{\vec{x}},\qquad \frac{d\vec{p}}{dt}=-m\vec{\nabla}\varphi
\label{4}
\end{equation}%
where the peculiar velocity $\vec{v}=a\dot{\vec{x}}$ relates to the momentum
as $\vec{v}=\vec{p}/ma$ and by virtue of (\ref{4}) obeys the equation
\begin{equation*}
\frac{d\vec{v}}{dt}+\frac{\dot{a}}{a}\vec{v}=-\frac{1}{a}\vec{\nabla}\varphi
.
\end{equation*}%
E.g., if $\vec{\nabla}\varphi =0$, then $\vec{p}=const$ and the velocity
changes as $\vec{v}\sim \frac{\vec{p}}{ma}$, i.e., $\sim 1/a\left( t\right) $%
.

Thus, to describe the cosmological model filled with a set of particles and
wormholes we get the system of equations as follows:
\begin{equation*}
\ddot{a}=-\frac{4}{3}\pi G\left( \rho _{b}(t)+\frac{3P_{b}(t)}{c^{2}}\right)
a
\end{equation*}%
\begin{equation}
\Delta \varphi =4\pi G\left( \delta \rho (\vec{x},t)+\frac{3\delta P(\vec{x}%
,t)}{c^{2}}\right) a^{2}  \label{PS}
\end{equation}%
\begin{equation*}
m_{A}a^{2}\dot{\vec{x}}_{A}=p_{A}
\end{equation*}%
\begin{equation*}
\dot{\vec{p}}_{A}=-m_{A}\vec{\nabla}\varphi (x_{A})
\end{equation*}%
where $A=1,...,N$ numerates particles or wormholes.

This system of equations looks as the ordinary one but we should take into
account that wormholes insert some complex (depending on time)
identification of points (boundary conditions for $\varphi $ and $\vec{x}_{A}
$). Moreover, since wormholes are described by a larger number of parameters
than a particle, i.e., $\eta _{A}$, $\dot{\eta}_{A}$ ($\eta =\left(
R_{+},R_{-},b,U_{\beta }^{\alpha }\right) $) to complete this system we
should add equations which govern the evolution of the rest parameters ($%
b_{A}$ and $U_{\beta A}^{\alpha }$). For the sake of simplicity in the
present paper we shall neglect the possible evolution of $b$ and $U_{\beta
}^{\alpha }$.

\section{ Boltzmann -Vlasov type equations\label{KD}}

In what follows we shall use the standard methods (e.g., see \cite{kin}).
Let us introduce the density of particles in the phase space as $\Gamma
_{m}=(\vec{x},\vec{p})$,
\begin{equation*}
dN_{m}=f_{m}(\Gamma _{m},t)d\Gamma _{m},~d\Gamma _{m}=d^{3}xd^{3}p
\end{equation*}%
and the number density of wormholes
\begin{equation*}
dN_{w}\left( \gamma \right) =F_{w}(\Gamma _{w},\gamma ,t)d\Gamma _{w}.
\end{equation*}%
here, since we assume the part of parameters $\gamma =(b,U_{\beta }^{\alpha
})$ be fixed (non-dynamical ones), $d\Gamma _{w}=d\eta dp_{\eta }$, $\Gamma
_{w}=(\eta ,p_{\eta })$, and, $\eta =(\vec{R}_{+},\vec{R}_{-})$. We also
note that according to (\ref{vr}) only one momentum is free. It is
convenient to consider both momenta $p_{\eta }=(P_{+},P_{-})$ as
independent, while the relation (\ref{vr}) is accounted for by a
delta-function - type multiplier in $F_{w}$. Then the matter density can be
expressed via $f_{m}(\vec{x},\vec{p},t)$ as
\begin{equation}
\rho _{m}(\vec{x},t)=\frac{m}{a^{3}}\int \delta \left( \vec{x}-\vec{x}%
^{\prime }\right) f_{m}(\Gamma _{m}^{\prime },t)d\Gamma _{m}^{\prime }=\rho
_{b}(t)\left[ 1+\delta _{m}(\vec{x},t)\right]  \label{dm}
\end{equation}%
and analogously the contribution from wormholes is (here $M\left( \gamma
\right) =\frac{4}{3}\pi b^{3}a^{3}\rho _{b}$)
\begin{equation*}
\rho _{w}(\vec{x},t)= \sum_{\sigma =\pm }\int \frac{M\left( \gamma \right) }{%
a^{3}} \delta \left( \vec{x}-\vec{R}_{\sigma}\right) F_{w}(\Gamma
_{w},\gamma ,t)d\Gamma _{w}d\gamma .
\end{equation*}%
We point out that in the proper (expanding) reference system the mean number
of particles in the volume element $\left( \Delta x\right) ^{3}$ remains
constant, i.e., $a^{3}\rho _{b}\sim const$ and so do the masses of wormholes
$M\left( \gamma \right) =M\left( b\right) \sim const$.

In $F_{w}(\Gamma _{w},\gamma ,t)$ indexes $w$ and $\gamma $ numerate merely
the sort of wormholes. It is convenient to introduce the reduced
distribution function for wormhole throats as follows%
\begin{equation*}
f_{w}\left( \vec{X},\vec{P},\gamma ,t\right) = \sum_{\sigma =\pm }\int
\delta \left( \vec{X}-\vec{R}_{\sigma }\right) \delta \left( \vec{P}-\vec{P}%
_{\sigma }\right) F_{w}(\Gamma _{w},t)d\Gamma _{w}
\end{equation*}%
which defines the mass density in the form analogous that of particles%
\begin{equation}
\rho _{w}(\vec{x},t)=\int \frac{M( \gamma )}{a^{3}} f_{w}\left( \vec{x},\vec{%
P},\gamma ,t\right) d^{3}\vec{P}d\gamma =\rho _{b}(t)\left[ 1+\delta _{w}(%
\vec{x},t)\right] .  \label{dw}
\end{equation}%
Then the system of Boltzmann-Vlasov equations are eqs. (\ref{dm}), (\ref{dw}%
),
\begin{equation}
\frac{\partial f_{A}}{\partial t}+\frac{\vec{p}}{m_{A}a^{2}}\nabla
f_{A}-m_{A}\nabla \varphi \frac{\partial f_{A}}{\partial \vec{p}}%
=\sum_{B}stf_{AB},  \label{5}
\end{equation}%
and the generalized Poisson equation (we set $\delta P/c^{2}=\frac{c_{s}^{2}%
}{c^{2}}\delta \rho \ll \delta \rho $ otherwise one has to add the
coefficient ($1+3c_{s}^{2}/c^{2}$) before $\delta \rho $)
\begin{equation}
\frac{1}{a^{2}}\Delta \varphi =4\pi G\left( \delta \rho (\vec{x},t)+\int
B\left( x,x^{\prime }\right) \delta \rho ^{\prime }d^{3}\vec{x}^{\prime
}\right) .  \label{PSN}
\end{equation}%
In the last equation the bias $B\left( x,x^{\prime }\right) $ accounts for
the topological polarizability of space in the presence of the gas of
wormholes (i.e., the proper boundary conditions at wormhole throats). As it
was shown by \cite{KS07} it expresses completely via $F_{w}(\Gamma
_{w},\gamma ,t)$ and is given by (\ref{bk0}), (\ref{bk1}). In (\ref{5}) the
index $A$ denotes either $m$ in the case of particles, or $(w,\gamma )$ in
the case of wormhole throats.

Rigorously speaking the above system is not complete, since the collisions
between particles and a wormhole involve both throats
\begin{equation}
stf_{AB}=\int w\left( \Gamma _{A}^{\prime },\Gamma _{B}^{\prime },\Gamma
_{A},\Gamma _{B}\right) \left[ f_{A}^{\prime }F_{B}^{\prime }-f_{A}F_{B}%
\right] d\Gamma _{A}^{\prime }d\Gamma _{B}^{\prime }d\Gamma _{B}  \label{sct}
\end{equation}%
where the scattering matrix $w\left( \Gamma _{A}^{\prime },\Gamma
_{B}^{\prime },\Gamma _{A},\Gamma _{B}\right) $ stands for the
transformation law $\left( \Gamma _{A},\Gamma _{B}\right) \rightarrow \left(
\Gamma _{A}^{\prime },\Gamma _{B}^{\prime }\right) $ which is defined by (%
\ref{tr1})-(\ref{tr3}). The function $F_{A}$ obeys the equation similar to
the two-point correlation function
\begin{equation}
\frac{\partial F_{A}}{\partial t}+\sum_{\sigma =\pm }\left( \frac{\vec{p}%
_{\sigma }}{m_{A}a^{2}}\vec{\nabla}_{\sigma }-m_{A}\vec{\nabla}_{\sigma
}\varphi \frac{\partial }{\partial \vec{p}_{\sigma }}\right)
F_{A}=\sum_{B}stF_{AB}.  \label{TP}
\end{equation}%
Then equations (\ref{5}) are obtained by the integration of the last
equation over the extra variables.

\section{Fluid equations\label{FD}}

Upon integrating (\ref{5}) over $\int $ $d^{3}p$ we get
\begin{equation}
\frac{\partial }{\partial t}\frac{\rho _{A}a^{3}}{m_{A}}+\frac{1}{a}\vec{%
\nabla}\left( \frac{\rho _{A}a^{3}}{m_{A}}\vec{u}_{A}\right) =D_{A},
\label{7}
\end{equation}%
which represents the first hydrodynamic equation (the discontinuity
equation)). The term $D_{A}=\int stf_{A}d^{3}p$ we define latter on, while
we used the notions for the density
\begin{equation*}
\int f_{A}d^{3}p=\frac{\rho _{A}(\vec{x},t)a^{3}}{m_{A}},
\end{equation*}%
the peculiar macroscopic velocity $\vec{u}(\vec{x},t)$
\begin{equation}
\vec{u}_{A}(\vec{x},t)=\frac{1}{m_{A}a}\frac{\int \vec{p}f_{A}d^{3}\vec{p}}{%
\int f_{A}d^{3}\vec{p}},  \label{6a}
\end{equation}%
and $\rho _{A}(\vec{x},t)=\rho _{b}(t)(1+\delta _{A}(\vec{x},t))$. Due to
the property $\rho _{b}a^{3}=const$ this equation can be rewritten as
\begin{equation}
\frac{\partial }{\partial t}\delta _{A}(\vec{x},t)+\frac{1}{a}\nabla _{\beta
}\left( \left( 1+\delta _{A}\right) u_{A}^{\beta }\right) =\frac{m_{A}}{\rho
_{b}a^{3}}D_{A}.  \label{7a}
\end{equation}%
We point out that in the standard hydrodynamics the terms of the type $D_{A}$
vanish due to the local conservation laws (of the number of particles,
momentum, energy). In collisions of particles and wormholes such terms
retain. During the scattering of particles on wormholes the number of
throats conserves locally which gives $D_{w}=0$. Though, in general, there
remain processes of throats merging which produce some value $D_{w}<0$. It
is also easy to see that all such terms vanish in the absence of
perturbations (for the homogeneous background).

Multiplying the equation (\ref{5}) with $p_{\alpha }$ and integrating over
momenta $d^{3}p$ we get the second (Eiller) equation
\begin{equation}
\frac{\partial }{\partial t}\left( a^{4}\rho _{A}u_{A}^{\alpha }\right)
+\rho _{A}a^{3}\nabla _{\alpha }\varphi +a^{3}\nabla _{\beta }\left(
u_{A}^{\alpha }u_{A}^{\beta }\rho _{A}\right) +a^{3}\nabla _{\alpha
}P=Q_{A}^{\alpha }  \label{EE}
\end{equation}%
where
\begin{equation*}
Q_{A}^{\alpha }=\int p^{\alpha }stf_{A}d^{3}p
\end{equation*}%
and we used the obvious definitions
\begin{equation}
\langle v^{\alpha }v^{\beta }\rangle =\frac{\int p^{\alpha }p^{\beta }fd^{3}p%
}{m^{2}a^{2}\int fd^{3}p},  \label{8}
\end{equation}%
\begin{equation*}
\langle v^{\alpha }v^{\beta }\rangle =u_{A}^{\alpha }u_{A}^{\beta }+\frac{1}{%
3}\delta _{\alpha \beta }\langle \delta v^{2}\rangle =u_{A}^{\alpha
}u_{A}^{\beta }+\frac{P_{A}}{\rho _{A}}\delta _{\alpha \beta },
\end{equation*}%
where $P_{A}$ is the pressure. The system of equations (\ref{7}) and (\ref%
{EE}) represents the basic hydrodynamic equations. In the same way one can
also add to the above system the equation for the heat transport (i.e., the
next momentum for the internal energy $m_{A}\langle \delta v^{2}\rangle /2$%
). For the sake of simplicity we do not consider the heat equation here. For
adiabatic linear perturbations such equation is not important. It however
becomes important more latter on non-linear stages of the formation of
astrophysical objects.

In view of $\rho _{b}a^{3}=const$ (\ref{EE}) maybe rewritten as ($H=\dot{a}%
/a $ is the Hubble constant)%
\begin{equation*}
\left( \frac{\partial }{\partial t}+H\right) \left( \left( 1+\delta
_{A}\right) u_{A}^{\alpha }\right) +\frac{1}{a}\left( 1+\delta _{A}\right)
\nabla _{\alpha }\varphi +
\end{equation*}%
\begin{equation*}
+\frac{1}{a}\nabla _{\beta }\left( \left( 1+\delta _{A}\right) u_{A}^{\alpha
}u_{A}^{\beta }\right) +\frac{1}{a\rho _{b}}\nabla _{\alpha }P_{A}=\frac{1}{%
a^{4}\rho _{b}}Q_{A}^{\alpha }.
\end{equation*}%
Let us take the divergency from the last equation and in view of (\ref{7a})
we find the master equations which govern the evolution of density
perturbations (here $Q_{A}=\nabla _{\alpha }Q_{A}^{\alpha }$)%
\begin{equation*}
\left( \frac{\partial }{\partial t}+2H\right) \left( \frac{\partial }{%
\partial t}\delta _{A}-\frac{m}{\rho _{b}a^{3}}D_{A}\right) =\frac{1}{a^{2}}%
\nabla _{\alpha }\left( 1+\delta _{A}\right) \nabla _{\alpha }\varphi +
\end{equation*}%
\begin{equation*}
+\frac{1}{a^{2}}\nabla _{\alpha }\nabla _{\beta }\left( \left( 1+\delta
_{A}\right) u_{A}^{\alpha }u_{A}^{\beta }\right) +\frac{1}{a^{2}\rho _{b}}%
\nabla ^{2}P_{A}-\frac{1}{a^{5}\rho _{b}}Q_{A}.
\end{equation*}%
Retaining just linear terms we get the equation for linear perturbations as
\begin{equation}
\left( \frac{\partial }{\partial t}+2H\right) \left( \frac{\partial }{%
\partial t}\delta _{A}-\frac{m}{\rho _{b}a^{3}}D_{A}\right) =\frac{1}{a^{2}}%
\Delta \left( \varphi +c_{A,s}^{2}\delta _{A}\right) -\frac{1}{a^{5}\rho _{b}%
}Q_{A}  \label{Le}
\end{equation}%
where we used the relation $\delta P=c_{s}^{2}\delta \rho $ and $c_{s}$ is
the sound speed. We also have to add the Poisson equation
\begin{equation}
\frac{1}{a^{2}}\Delta \varphi =4\pi G\rho _{b}\sum_{A}\left( \delta
_{A}+\int B\left( x,x^{\prime }\right) \delta _{A}^{\prime }d^{3}\vec{x}%
^{\prime }\right) .  \label{Lp}
\end{equation}

Thus, the system (\ref{Le}) and (\ref{Lp}) represents the master equation
which describes the development of linear adiabatic perturbations in the
presence of the gas of wormholes. This system represents the standard system
for CDM model where wormholes play the role of dark matter particles. The
difference appears however due to the additional terms $D_{A}$, $Q_{A}$, and
the bias $B\left( x,x^{\prime }\right) $. For homogeneous background some
terms disappear. As we shall see latter on they disappear in the long-wave
limit as well i.e., as $k\ll k_{J}$ (where $k_{J}$ is the Jeans wavelength
whose definition in the presence of wormholes somewhat differs from the
standard one).

\section{Kinetic coefficients $D$ and $Q$}

Consider an arbitrary point $\vec{x}$ on the sphere $S_{-}$, i.e., $\vec{x}%
\in S_{-}$ and therefore $\xi _{-}^{2}=$ $\left( \vec{x}-\vec{R}_{-}\right)
^{2}$ $=$ $b_{w}^{2}$. The gluing procedure transforms this point into a
conjugated point $\vec{x}^{\prime }\in S_{+}$ which has the form $\vec{x}%
^{\prime }=\vec{R}_{+}+\vec{\xi}_{+}$ where $\vec{\xi}_{+}$ relates to $\vec{%
\xi}_{-}$ by some rotation $\xi _{+}^{\alpha }=U_{\beta }^{\alpha }\xi
_{-}^{\beta }$. Let $\Gamma _{w}=\left( R_{\pm },b_{w},U,V_{\pm }\right) $
denote the set of parameters of the wormhole. The scattering matrix $W$ is
presented as $W=W_{+}+W_{-}$ where $W_{\pm }$ corresponds to which of
throats absorbs particles. Due to the obvious symmetry between throats this
gives merely the factor $2$ in final expressions. Then we find from (\ref%
{tr1})-(\ref{tr3}) $W_{\pm }(\Gamma _{m},\Gamma _{w},\Gamma _{m}^{\prime
},\Gamma _{w}^{\prime },)=\left\vert u_{\pm }\right\vert \sigma _{\pm
}(\Gamma _{m},\Gamma _{w},\Gamma _{m}^{\prime },\Gamma _{w}^{\prime },)$
where
\begin{equation}
\sigma _{-}=\delta \left( \xi _{+}-b_{w}\right) \delta \left( \Gamma
_{m}^{\prime }-\Gamma _{m,-}\right) \delta \left( \Gamma _{w}^{\prime }-%
\widetilde{\Gamma }_{w}\right) -  \label{w}
\end{equation}%
\begin{equation*}
-\delta \left( \xi _{-}-b_{w}\right) \delta \left( \Gamma _{m}^{\prime
}-\Gamma _{m}\right) \delta \left( \Gamma _{w}^{\prime }-\Gamma _{w}\right)
\end{equation*}%
and analogous term $\sigma _{+}$ with the obvious replacement ($-\rightarrow
+$ and $U\rightarrow U^{-1}$), where we used the notions as follows $\vec{\xi%
}_{\pm }=\vec{x}-\vec{R}_{\pm }$, $\vec{n}_{\pm }=\vec{\xi}_{\pm }/b_{w}$,
\begin{equation*}
\Gamma _{m,-}=\left( x_{-},p_{-}\right) ,~\ \widetilde{\Gamma }_{w}=\left(
R_{\pm },V_{\pm }^{\prime },b_{w},U\right) ,
\end{equation*}%
and the relations
\begin{equation*}
x_{-}^{\alpha }=R_{+}^{\alpha }+U_{\beta }^{\alpha }\xi _{-}^{\beta },
\end{equation*}%
\begin{equation}
\vec{p}_{-}=U\left( \vec{p}-\frac{2M_{w}ma}{M_{w}+m}\left( u_{-}n_{-}\right)
\vec{n}_{-}\right) ,  \label{trf}
\end{equation}

\begin{equation}
\vec{V}_{\pm }^{\prime }=\vec{V}_{\pm }+\frac{2m}{M_{w}+m}\left( u_{\pm
}n_{\pm }\right) \vec{n}_{\pm },\
\end{equation}%
and
\begin{equation*}
\vec{u}_{-}=\vec{v}-\vec{V}_{-},\ \ \ u_{+}^{\alpha }=U_{\beta }^{\alpha
}u_{-}^{\beta }.
\end{equation*}

For astrophysical needs wormholes may be considered as point-like objects.
This is achieved by the replacement in (\ref{w})%
\begin{equation}
\delta \left( \xi _{\pm }-b_{w}\right) \rightarrow 4\pi b_{w}^{2}\delta
\left( \vec{R}_{\pm }-\vec{x}\right)  \label{delta}
\end{equation}%
where $\delta \left( \vec{x}\right) $ is the 3-dimensional delta function $%
\int \delta \left( \vec{x}\right) d^{3}x=1$. Thus the above expressions
completely define the scattering matrix $W$ and the scattering term
\begin{equation}
stf_{AB}=\int W\left( \Gamma _{A}^{\prime },\Gamma _{B}^{\prime },\Gamma
_{A},\Gamma _{B}\right) \left[ f_{A}^{\prime }F_{B}^{\prime }-f_{A}F_{B}%
\right] d\Gamma _{A}^{\prime }d\Gamma _{B}^{\prime }d\Gamma _{B}.
\end{equation}

\subsection{The scattering of particles}

Consider first the terms $D_{m}$ and $Q_{m}$. Since we suppose that $%
M_{w}\gg m$, it follows that $\left\langle V_{\pm }^{2}\right\rangle \sim
T/\left( aM_{w}\right) \ll \left\langle v^{2}\right\rangle \sim T/\left(
am\right) $ and to the leading order we can neglect the motions of wormholes
which essentially simplifies the above expressions (e.g., $u_{-}=v=p/ma$ in (%
\ref{w}) and (\ref{trf})). We also suppose the homogeneous distribution of
wormholes in space which gives $f_{w}\left( \eta \right) =\int f_{w}\left(
\Gamma _{w}\right) dP_{\eta }=F\left( \left\vert \vec{R} _{-}-\vec{R}%
_{+}\right\vert ,\gamma \right) $ (here $\gamma = b_w, U$). Then for the
scattering term (\ref{sct}) we find
\begin{equation*}
stf_{{m,w}}=8\pi \int \left\vert v\right\vert \left[ f_{m}^{\prime }\left(
R,p_{-}\right) -f_{m}\left( x,p\right) \right] b_{w}^{2}F\left( \left\vert
x-R\right\vert ,\gamma \right) d\mu ,
\end{equation*}%
where $d\mu=d^3pd^{3}Rd\gamma$. This defines the kinetic coefficient $D_{m}$
for particles as (we use here the property $\left\vert v\right\vert
=\left\vert v^{\prime }\right\vert $ and $d^{3}p=d^{3}p^{\prime }$
\begin{equation*}
D_{m}=\frac{8\pi }{ma}\int \left[ \left\vert p^{\prime }\right\vert
f_{m}^{\prime }\left( R,p^{\prime }\right) -\left\vert p\right\vert
f_{m}\left( x,p\right) \right] g\left( \left\vert x-R\right\vert \right)
d^{3}Rd^{3}p,
\end{equation*}%
where $g\left( R\right) =\int b_{w}^{2}F\left( R,\gamma \right) d\gamma $
which gives%
\begin{equation}
D_{m}=\frac{8\pi }{ma}\int \left[ K\left( R+x\right) -K\left( x\right) %
\right] g\left( R\right) d^{3}R  \label{D}
\end{equation}%
with%
\begin{equation}
K\left( x,t\right) =\int \left\vert p\right\vert f\left( x,p,t\right) d^{3}p.
\label{K}
\end{equation}%
We point out that for the homogeneous distribution of particles $K\left(
x\right) =K=const$ and $D_{m}\equiv 0$.

Consider now the second coefficient $Q_{m}^{\alpha }$ which is given by
\begin{equation}
Q_{m}^{\alpha }=\frac{8\pi }{ma}\int \left[ \frac{1}{9}K_{\alpha }\left(
R+x\right) -K_{\alpha }\left( x\right) \right] g\left( R\right) d^{3}R,
\label{Ba}
\end{equation}%
where
\begin{equation}
K_{\alpha }\left( x\right) =\int p_{\alpha }\left\vert p\right\vert f\left(
x,p\right) d^{3}p.  \label{Ka}
\end{equation}%
The multiplier $1/9$ appears in (\ref{Ba}) upon transformations as follows.
According to (\ref{trf})
\begin{equation*}
K_{\alpha }^{\prime }\left( x\right) =\int \left[ p_{\alpha }\left\vert
p^{\prime }\right\vert f\left( x,p^{\prime }\right) \right] d^{3}p^{\prime }=
\end{equation*}%
\begin{equation*}
=U_{\alpha \beta }\int \left( p_{\beta }^{\prime }-2\left( p^{\prime
}n\right) n_{\beta }\right) \left\vert p^{\prime }\right\vert f\left(
x,p^{\prime }\right) d^{3}p^{\prime }.
\end{equation*}%
The replacement (\ref{delta}) means an additional averaging over $n_{\beta }$
which gives $<n_{\alpha }n_{\beta }>=\frac{1}{3}\delta _{\alpha }{}_{\beta }$%
, while assuming an isotropic distribution of wormholes over the rotation
matrix $U$ gives an additional multiplier $<U_{\alpha \beta }>=\frac{1}{3}%
\delta _{\alpha }{}_{\beta }$. Thus, we see that $K_{\alpha }^{\prime
}\left( x\right) =\frac{1}{9}K_{\alpha }\left( x\right) $ in (\ref{Ba}). The
above expressions (\ref{D}) and (\ref{Ba}) show that the kinetic
coefficients are expressed via the momenta $K\left( x\right) $ and $%
K_{\alpha }\left( x\right) $.

\subsection{functions $K(x) $ and $K_{\protect\alpha }(x) $}

Keeping in mind the linearized equations (\ref{Le}) we will use the
following form for the quasi-equilibrium distribution function%
\begin{equation*}
f_{0}\left( x,p\right) =\frac{n( x,t) }{\left( 2\pi maT\left( x,t\right)
\right) ^{3/2}}\exp \left( -\frac{\left( p-mau\left( x,t\right) \right) ^{2}%
}{2maT\left( x,t\right) }\right)
\end{equation*}%
which corresponds to the locally thermodynamic equilibrium distribution and $%
n\left( x,t\right) =\rho (\vec{x},t)a^{3}/m$. Then we find
\begin{equation*}
K\left( n,u,T\right) =\frac{n}{( 2\pi maT) ^{3/2}}\int \exp \left( -\frac{%
\left( \vec{p}-ma\vec{u}\right) ^{2}}{2maT}\right) \left\vert p\right\vert
d^{3}p.
\end{equation*}%
Expanding this by $\vec{u}$ (i.e., $K=K^{0}+K_{\alpha }^{1}u_{\alpha }+\frac{%
1}{2}K_{\alpha \beta }^{2}u_{\alpha }u_{\beta }+...,$) we find
\begin{equation}
K=n\frac{2}{\sqrt{\pi }}\left( 2maT\right) ^{1/2}\left( 1+\frac{ma}{6T}%
u^{2}\right) +....  \label{K0}
\end{equation}%
To evaluate the function $K_{\alpha }\left( x\right) $ we point out the
obvious relations%
\begin{equation*}
\frac{\partial }{\partial u_{\alpha }}f_{0}\left( x,p\right) =-\frac{%
mau_{\alpha }-p_{\alpha }}{T}f_{0}\left( x,p\right)
\end{equation*}%
which defines the function $K_{\alpha }\left( x\right) $ in the form
\begin{equation*}
K_{\alpha }\left( x\right) =T\left( \frac{\partial }{\partial u_{\alpha }}+%
\frac{mau_{\alpha }}{T}\right) K
\end{equation*}%
and from (\ref{K0}) we find
\begin{equation*}
K_{\alpha }\left( x\right) \simeq 2Tn\sqrt{\frac{2maT}{\pi }} \left( \frac{%
\partial }{\partial u_{\alpha }}+\frac{mau_{\alpha }}{T}\right) \left( 1+%
\frac{ma}{6T}u^{2}\right).
\end{equation*}%
Thus, in the leading order by $\vec{u}$ we get%
\begin{equation*}
K \simeq \frac{2n}{\sqrt{\pi }}(2maT) ^{1/2},\ \ K_{\alpha }\left(
x,t\right) \simeq \frac{4}{3}K\left( x,t\right) mau^{\alpha }\left(
x,t\right) .
\end{equation*}%
Using the relations%
\begin{equation*}
\langle \delta v^{2}\rangle =\frac{3P(\vec{x},t)}{\rho (\vec{x},t)}=\frac{1}{%
m^{2}a^{2}}\frac{\int p^{2}fd^{3}p}{\int fd^{3}p}=\frac{3maT}{m^{2}a^{2}}
\end{equation*}%
or $T=maP/\rho $, we may rewrite the function $K$ in the equivalent form
\begin{equation*}
K\left( x,t\right) \simeq \frac{2\sqrt{2}}{\sqrt{\pi }}\rho a^{4}c_{s}\left(
\frac{P}{\rho c_{s}^{2}}\right) ^{1/2} .
\end{equation*}

For linear perturbations we shall use $\delta P_{m}=c_{s}^{2}\delta \rho
_{m}=c_{s}^{2}\rho _{b}(t)\delta _{m}$ which gives
\begin{equation}
\delta K\left( x,t\right) \simeq \frac{1}{2}K\left( t\right) \left( \frac{%
\rho }{P}c_{s}^{2}+1\right) \frac{\delta \rho }{\rho },\ \   \label{Kt}
\end{equation}%
and, therefore, we get the functions we are looking for in the form%
\begin{equation}
\delta K\simeq \frac{1}{2}K\left( t\right) \left( 1+c_{s}^{2}\frac{\rho _{b}%
}{P}\right) \delta (\vec{x},t),\ \ \ \delta K_{\alpha }\simeq \frac{4}{3}%
K\left( t\right) mau_{\alpha }\left( x,t\right) .  \label{Kl}
\end{equation}

\subsection{Kinetic coefficients $D_{m}$ and $Q_{m}$}

In what follows we shall use the parameter $P_{m}/\rho _{m}c_{s}^{2}\simeq 1$
which gives $\delta K\simeq K(t) \delta (\vec{x},t)$ and $K \simeq \frac{2%
\sqrt{2}}{\sqrt{\pi }}\rho a^{4}c_{s}$. Then by the use of the above
formulas we finally find from (\ref{D}) the kinetic coefficients as%
\begin{equation}
D_{m}\left( x,t\right) =\frac{8\pi }{ma}K \int \left[ \delta _{m}\left(
x+R,t\right) -\delta _{m}\left( x,t\right) \right] g\left( R\right) d^{3}R
\end{equation}%
and analogously from (\ref{Ba})
\begin{equation}
Q_{m}=\frac{32\pi }{3}K \int \left[ \frac{1}{9}\nabla _{\alpha
}u_{m}^{\alpha }\left( x+R,t\right) -\nabla _{\alpha }u_{m}^{\alpha }( x,t) %
\right] g\left( R\right) d^{3}R.
\end{equation}%
In the last equation the term $\nabla _{\alpha }u_{m}^{\alpha }$ can be
expressed from (\ref{7a}) as
\begin{equation*}
\nabla _{\alpha }u_{m}^{\alpha }=-a\left( \frac{\partial }{\partial t}\delta
_{m}-\frac{m }{\rho _{b}a^{3}}D_{m}\right),
\end{equation*}
while for the quantity $\frac{m }{\rho _{b}a^{3}}D_{m}$ we find
\begin{equation*}
\frac{m}{\rho _{b}a^{3}}D_{m}\simeq 16\sqrt{2\pi }c_{s}\int \left[ \delta
\left( x+R,t\right) -\delta \left( x,t\right) \right] g\left( R\right)
d^{3}R.
\end{equation*}

Consider now some qualitative estimates. Let $\int g\left( R\right)
d^{3}R=<b_{w}^{2}>n_{w}$ ($n_{w}$ is the density of wormholes in the
commoving reference system, the physical density is $n_{w}/a^{3}$). Then we
may estimate%
\begin{equation*}
\frac{m}{\rho _{b}a^{3}}D_{m}\sim 16\sqrt{2\pi }\frac{1}{\tau _{w}}\frac{%
d_{w}}{L}\delta \sim \nu _{w}\delta
\end{equation*}%
where $\tau _{w}\sim \left( c_{s}b_{w}^{2}n_{w}\right) ^{-1}$, $d$ is the
characteristic distance between throats, and $L$ is the characteristic scale
of the inhomogeneity. For estimates we also can suppose $n_{w}=\rho
_{w}a^{3}/M_{w}$, where $M_{w}=\rho _{b}\frac{4}{3}\pi a^{3}b_{w}^{3}$ which
gives
\begin{equation*}
\tau _{w}\sim \frac{M_{w}}{c_{s}b_{w}^{2}\rho _{w}a^{3}}=\frac{4}{3}\pi
\frac{b_{w}}{c_{s}}\frac{\rho _{b}}{\rho _{w}}
\end{equation*}%
and finally we define the estimate for the collision frequency
\begin{equation*}
\nu _{w}\left( k\right) \sim \frac{6\sqrt{2}}{\pi \sqrt{\pi }}\frac{d_{w}}{%
b_{w}}\frac{\rho _{w}}{\rho _{b}}kc_{s}.
\end{equation*}%
It can be seen that for $kc_{s}\rightarrow 0$ this correction is negligible
(the collision frequency $\nu _{w}\rightarrow 0$), while for rather short
wave-length this term may give the leading contribution.

\subsection{Kinetic coefficients $D_{w}$ and $Q_{w}$}

As it was already pointed out the collisions conserve the local number of
throats which immediately gives the value
\begin{equation}
D_{w}=0.  \label{DW}
\end{equation}%
Some non-vanishing value of $D_{w}<0$ may appear when we take into account
for the processes of throats merging (annihilation of wormholes). Such
processes are important on the very early stages of the evolution of the
Universe, since they are responsible for the formation of the background
distribution of wormholes in space. E.g., the origin of the Tully-Fisher
relation for spirals was suggested by \cite{KT06} which requires the decay
for some portion of primordial wormholes. Such a decay is accompanied with
an essential reheating, since every merging of a wormhole throats radiates
the energy $\sim M_{w}c^{2}$. We however leave this problem aside for the
future research.

In the case of the isotropic background the second coefficient $Q_{w}$ does
not require a separate evaluation. In spite of the fact that during the
scattering the momentum does not conserves (\ref{cm}), some kind of the
conservation law for mean values takes however place. Indeed, due to the
obvious symmetries $S_{+}^{3}\leftrightarrow S_{-}^{3}$ and $t\rightarrow -t$%
\begin{equation*}
w\left( \Gamma _{A}^{\prime },\Gamma _{B}^{\prime },\Gamma _{A},\Gamma
_{B}\right) =w\left( \Gamma _{A},\Gamma _{B},\Gamma _{A}^{\prime },\Gamma
_{B}^{\prime }\right)
\end{equation*}%
we find that%
\begin{equation*}
Q_{m,\alpha }=\left\langle p_{\alpha }\right\rangle =\frac{1}{2}\left\langle
p_{\alpha }-p_{\alpha }^{\prime }\right\rangle ,
\end{equation*}%
where $\left\langle p_{\alpha }\right\rangle =\int p_{\alpha
}stf_{m,w}d^{3}p $. And analogous expression holds for throats $Q_{w,\alpha
}=\left\langle P_{\pm }\right\rangle $ $=$ $\frac{1}{2}\left\langle P_{\pm
}-P_{\pm }^{\prime }\right\rangle $ $=$ $\int P_{\pm \alpha
}stf_{w,m}d^{3}P_{\pm }$. Due to the symmetry $S_{+}^{3}\rightarrow
S_{-}^{3} $ we can always change $\left\langle P_{+}\right\rangle
=\left\langle P_{-}\right\rangle $. Then using the relation (\ref{cm}) we
find%
\begin{equation*}
\frac{1}{2}\left\langle p-p^{\prime }\right\rangle +\frac{1}{2}\left\langle
P_{-}-P_{+}^{\prime }\right\rangle =\frac{1}{2}\left\langle \left(
1-U\right) p\right\rangle +\frac{1}{2}\left\langle \left( 1-U\right)
P_{-}\right\rangle .
\end{equation*}%
For the isotropic distribution the averaging over the matrix $U$ gives $%
<U_{\alpha \beta }>=\frac{1}{3}\delta _{\alpha }{}_{\beta }$ and, therefore,
the above expression transforms to $Q_{m}=-Q_{w}+\frac{1}{3}Q_{m}+\frac{1}{3}%
Q_{w}$, i.e.,
\begin{equation}
Q_{m}=-Q_{w}.  \label{QW}
\end{equation}%
The last equation means not more than the conservation of the total mean
momentum density. It expresses the balance for the transformation of the
momentum between particles and wormholes.

\section{The behavior of linear perturbations\label{PR}}

Consider now the Fourier transform for the density perturbations%
\begin{equation*}
\delta _{A}\left( x\right) =\left( 2\pi \right) ^{-3/2}\int e^{i\vec{k}\vec{x%
}}\delta _{A,\vec{k}}d^{3}k.
\end{equation*}%
Then we find
\begin{equation}
\frac{m}{\rho _{b}a^{3}}D_{A}\left( k\right) =-\nu _{A}\left( k\right)
\delta _{A,\vec{k}}  \label{DF}
\end{equation}%
where%
\begin{equation}
\nu _{m}\left( k\right) =\left( 8\pi \right) ^{2}c_{s}\left[ g\left(
0\right) -g\left( -\vec{k}\right) \right] ,\ \ \nu _{w}\left( k\right) =0
\end{equation}%
Analogously we find
\begin{equation}
\frac{1}{a^{5}\rho _{b}}Q_{m}\left( k\right) =-\frac{1}{a^{5}\rho _{b}}%
Q_{w}\left( k\right) =\Omega _{k}\left( \frac{\partial }{\partial t}+\nu
_{m}\left( k\right) \right) \delta _{m,\vec{k}}  \label{QWM}
\end{equation}%
where%
\begin{equation}
\Omega _{k}=\frac{4}{3}\left( 8\pi \right) ^{2}c_{s}\left[ g\left( 0\right) -%
\frac{1}{9}g\left( -\vec{k}\right) \right] .
\end{equation}%
Since we assume that $M_{w}\gg m$, with a very good approximation we may set
$\delta P_{w}=0$. Therefore, the master equations are
\begin{equation}
\left( \frac{\partial }{\partial t}+2H+\Omega _{k}\right) \left( \frac{%
\partial }{\partial t}+\nu _{m}\left( k\right) \right) \delta _{m,\vec{k}}+%
\frac{k^{2}c_{s}^{2}}{a^{2}}\delta _{m,\vec{k}}+\frac{k^{2}}{a^{2}}\varphi
_{k}=0  \label{master1}
\end{equation}%
\begin{equation}
\left( \frac{\partial }{\partial t}+2H\right) \frac{\partial }{\partial t}%
\delta _{w,\vec{k}}-\Omega _{k}\left( \frac{\partial }{\partial t}+\nu
_{m}\left( k\right) \right) \delta _{m,\vec{k}}+\frac{k^{2}}{a^{2}}\varphi
_{k}=0  \label{master2}
\end{equation}%
and the Poisson equation (for the Newton potential)
\begin{equation}
-k^{2}\frac{1}{a^{2}}\varphi _{k}=4\pi G\left( 1+B\left( k\right) \right)
\rho _{b}\left( \delta _{m,\vec{k}}+\delta _{w,\vec{k}}\right) ,
\label{PSN2}
\end{equation}%
where the function $B\left( k\right) $ was determined by \cite{KS07} and has
the form\footnote{%
We recall that the bias $B$ consists of the two terms (\ref{bk0}) and (\ref%
{bk1}). For perturbations the leading contribution comes from $B_{1}$, while
the first part $B_{0}$ is accounted for by the insertion of the rest masses
of wormholes.}
\begin{equation}
B\left( k\right) =\frac{8\pi \left( g_{1}\left( k\right) -g_{1}\left(
0\right) \right) }{k^{2}},  \label{b_k}
\end{equation}%
where $g_{1}( X) =\int b_{w}F( R,\gamma ) d\gamma $ (so that $\int
g_{1}\left( x\right) d^{3}x=<b_{w}>n_{w}$). We point out that all the
coefficients $\Omega _{k}$, $\nu _{m}\left( k\right) $, and $B\left(
k\right) $ are functions of the background (quasi-stationary) distribution
of wormholes $F\left( \left\vert R_{+}-R_{-}\right\vert ,\gamma \right) $
whose exact form requires an independent further investigation.

\subsection{Perturbations in the density of wormholes $\protect\delta _{w,%
\vec{k}}$}

Consider first the case when $\delta _{m,\vec{k}}=0$. Then the equation (\ref%
{master2}) with (\ref{PSN2}) taken into account reads%
\begin{equation*}
\left( \frac{\partial }{\partial t}+2H\right) \frac{\partial }{\partial t}%
\delta _{w,\vec{k}}-4\pi G_{k}\rho _{b}\delta _{w,\vec{k}}=0.
\end{equation*}%
where $G_{k}=G\left( 1+B\left( k\right) \right) $. We point out that this
equation coincides with the standard equation for perturbations in cold dark
matter particles. Some difference appears however due to the presence of the
bias $B( k )$ which reflects the polarizability of space filled with the gas
of wormholes. Formally such a polarizability looks like a scale-dependent
renormalization of the gravitational constant. By other words, at very large
scales (and on the linear stage of the development of perturbations) the gas
of wormholes reproduces exactly the dark matter particles.

The essential difference appears however on the non-linear stage due to the
existence of the mutual exchange with the momentum between wormholes and
baryons (\ref{QW}) (\ref{QWM}). It is clear that such an exchange cure the
basic failure of CDM (i.e., the presence of cusps in the center of
galaxies), since dark halos around galaxies should rotate.

\subsection{Matter perturbations $\protect\delta _{m,\vec{k}}$}

If we set $\delta _{w,\vec{k}}=0$, then the master equation for
perturbations becomes
\begin{equation}
\left( \frac{\partial }{\partial t}+2H+\Omega _{k}\right) \left( \frac{%
\partial }{\partial t}+\nu _{m}\left( k\right) \right) \delta +\left[ \frac{%
k^{2}c_{s}^{2}}{a^{2}}-4\pi G_k\rho _{b} \right] \delta =0.
\end{equation}%
The additional coefficients $\nu _{m}\left( k\right) $ and $\Omega _{k}$
have the clear physical interpretation. $\nu _{m}\left( k\right) $ is the
collision frequency which describes the processes of the absorption and
re-radiation of particles by wormholes. It is clear that such processes
somewhat smooth the initial inhomogeneity in the particle number density
which results in the specific damping we already discussed in \cite{KSZ}. At
very large distances $k\gg d$ (where $d$ is the characteristic distance
between throats) this kind of damping vanishes $\nu _{m}\left( k\right)
\rightarrow 0$. The additional friction term, proportional to $\Omega _{k}$,
describes the mutual interchange with momentum between particles and
wormholes. We point out that in the linear approximation there exists only a
transport of the momentum density from particles to wormholes which also
leads to an additional damping of inhomogeneities. Moreover, this kind of
damping retains in the long-wave limit as well ($\Omega _{0}\neq 0$).

Consider the redefinition
\begin{equation*}
\delta _{\vec{k}}=\exp \left( -\int^{t}\nu \left( k\right) dt\right)
\widetilde{\delta }_{\vec{k}}
\end{equation*}%
then, for the new quantity $\widetilde{\delta }_{\vec{k}}$ we find the
standard-type equation
\begin{equation}
\frac{\partial ^{2}}{\partial t^{2}}\widetilde{\delta }_{\vec{k}}+2H_{k}%
\frac{\partial }{\partial t}\widetilde{\delta }_{\vec{k}}+\left[ \frac{%
k^{2}c_{s}^{2}}{a^{2}}-4\pi G_{k}\rho _{b}\right] \widetilde{\delta }_{\vec{k%
}}=0  \label{delta1}
\end{equation}%
where
\begin{equation*}
2H_{k}=2H+\Omega _{k}-\nu \left( k\right) ,\text{ and }G_{k}=G\left(
1+B\left( k\right) \right) .
\end{equation*}%
By other words, this equation formally coincides with the standard one with
the renormalized friction $H_{k}$ and the gravitational $G_{k}$ constants.
The presence of the additional damping $\nu \left( k\right) $ and the
friction $\Omega _{k}$ somewhat weakens the standard Jeans instability.
Indeed, in the absence of the expansion ($H=0$) (\ref{delta1}) defines the
dispersion relations ($\delta _{\vec{k}}\sim \exp \left( \lambda t\right) $)
\begin{equation*}
(\lambda +\nu)^{2}+2H_{k}(\lambda +\nu)+\left[ \frac{k^{2}c_{s}^{2}}{a^{2}}%
-4\pi G_{k}\rho _{b}\right] =0
\end{equation*}%
which defines the two (decaying and increasing) modes $\delta _{\vec{k}%
}^{1,2}\sim \exp \left( \lambda _{1,2}t\right) $ as%
\begin{equation*}
\lambda _{1,2}=-\frac{\left( \Omega _{k}+\nu \left( k\right) \right) }{2}\pm
\sqrt{4\pi G_{k}\rho _{b}+\frac{\left( \Omega _{k}-\nu \left( k\right)
\right) ^{2}}{4}-\frac{k^{2}c_{s}^{2}}{a^{2}}}.
\end{equation*}%
In the limit $k\rightarrow 0$ we find for the rate of the growth
\begin{equation*}
\lambda _{1}\simeq \sqrt{4\pi G_{0}\rho _{b}+\frac{\Omega _{0}{}^{2}}{4}}-%
\frac{\Omega _{0}}{2}.
\end{equation*}

\subsection{Estimates for the background distribution of wormholes}

As it follows from (\ref{DF}) - (\ref{QWM}), (\ref{b_k}) the functions $\nu
\left( k\right) $, $\Omega _{k}$, and $B\left( k\right) $ depend on the
background distribution of wormholes in space. The background distribution
requires an independent consideration and we present it elsewhere, while in
this section we present some qualitative consideration.

As it can be seen from (\ref{TP}) the function $F_{w}$ obeys the equation
similar to the two-point correlation function. We consider here the two
different limits (the correct answer lies somewhere in the middle). First,
we consider the frozen wormholes (i.e., $V_{\pm }=0$). Then the function $%
F_{w}=F\left( \left\vert R_{-}-R_{+}\right\vert ,b_{w},U\right) $ is
completely determined by the process of the wormhole production during the
quantum stage (presumably an inlationary) of the evolution of the Universe.
Let us choose the simplest case when $F_{w}\sim \delta \left( \left\vert
R_{-}-R_{+}\right\vert -d_{w}\right) $ (e.g., see \cite{KS07}). Then we get%
\begin{equation*}
g\left( R\right) =\frac{\overline{b^{2}}_{w}n_{w}}{4\pi d_{w}^{2}}\delta
\left( R-d_{w}\right) ,\ \ g_{1}\left( R\right) =\frac{\overline{b}_{w}n_{w}%
}{4\pi d_{w}^{2}}\delta \left( R-d_{w}\right)
\end{equation*}%
where $n_{w}$is the density of throats, $\overline{b}_{w}$ is the mean
throat radius, and $d_{w}$ is the distance between throats. Therefore, we
find $g\left( k\right) $ $=$ $\overline{b^{2}}_{w}n_{w}(2\pi )^{-3/2}\sin
(kd_{w})/(kd_{w})$ which defines the functions $\nu \left( k\right) $, $%
\Omega _{k}$, and $B\left( k\right) $ in the form
\begin{equation*}
B\left( k\right) =-2n_{w}\overline{b}_{w}(2\pi )^{-1/2}\frac{1}{k^{2}}\left(
1-\frac{\sin \left( kd_{w}\right) }{kd_{w}}\right) ,
\end{equation*}%
\begin{equation*}
\nu _{m}\left( k\right) =\left( 8\pi \right) ^{2}c_{s}\overline{b^{2}}%
_{w}n_{w}(2\pi )^{-3/2}\left( 1-\frac{\sin \left( kd_{w}\right) }{kd_{w}}%
\right) ,
\end{equation*}%
\begin{equation*}
\Omega _{k}=\frac{4}{3}\left( 8\pi \right) ^{2}c_{s}\overline{b^{2}}%
_{w}n_{w}(2\pi )^{-3/2}\left[ 1-\frac{1}{9}\frac{\sin \left( kd_{w}\right) }{%
kd_{w}}\right] .
\end{equation*}%
which for $kd_{w}\ll 1$ give
\begin{equation*}
B\left( k\right) \approx -\frac{2n_{w}\overline{b}_{w}}{(2\pi )^{1/2}}\frac{1%
}{6}d_{w}^{2}(1-\frac{1}{20}\left( kd_{w}\right) ^{2}+...),
\end{equation*}%
\begin{equation*}
\nu _{m}\left( k\right) \approx \frac{\left( 8\pi \right) ^{2}}{6}c_{s}%
\overline{b^{2}}_{w}n_{w}(2\pi )^{-3/2}k^{2}d_{w}^{2},
\end{equation*}%
\begin{equation*}
\Omega _{k}\approx \frac{4}{3}\frac{\left( 8\pi \right) ^{2}}{9}c_{s}%
\overline{b^{2}}_{w}n_{w}(2\pi )^{-3/2}\left( 8+\frac{1}{6}%
k^{2}d_{w}^{2}+...\right) .
\end{equation*}%
Thus, we see that in the long-wave limit $kd_{w}\ll 1$ we have the behavior $%
\nu _{m}\sim k^{2}$, while $\Omega ,B\sim const$.

We point out that this case can presumably be far from reality, since the
presence of hot matter makes wormholes move chaotically. What we should
expect that $F_{w}\left( V\right) \sim \exp \left( -M_{w}V^{2}/2T\right) $
has the Maxwell-like form. However, due to the enormous typical value of the
rest mass $M_{w}\sim 10^{28}m_{p}$, the approximation of frozen wormholes
can in turn work sufficiently well.

The second case corresponds to the opposite situation, when all correlations
between conjugated throats are lost, i.e., when $F_{w}=F\left( \left\vert
R_{-}-R_{+}\right\vert \right) \sim const$. Then we find $g\left( k\right)
\sim g_{0}\delta \left( k\right) $ and all quantities $\nu \left( k\right) $%
, $\Omega _{k}$, and $B\left( k\right) $ tend to their asymptotic (as $%
k\rightarrow 0$) values.

\section{Summary}

In the present section we collect basic results. First of all we showed that
the background density generates a non-vanishing rest mass of a wormhole (%
\ref{M}). Therefore, wormholes may play the role of dark matter particles in
CDM models. Then, we considered the scattering between a particle and a
wormhole which straightforwardly shows that wormholes move in space. We
derived basic equations (\ref{tr1})-(\ref{tr3}) in the non-relativistic case
($V\ll c$) which however admit straightforward generalization to the
relativistic case. Based on the scattering equations we suggested the
kinetic description in Sec. \ref{KD} and derived the basic fluid equations
in Sec. \ref{FD}. All those equations are generalized straightforwardly to
the relativistic case as well and we present it elsewhere. In Sec. \ref{PR}
we derived equations for linear density perturbations and have shown that at
large scales density perturbations, related to wormholes, behave exactly
like standard dark matter particles. However, even on the linear stage there
always exists the transport of the momentum density from baryons to
wormholes. Already such a phenomenon is enough to cure the basic failure of
CDM particles, namely, to remove the cusps in centers of galaxies. Thus, we
suppose that at present wormholes represent the best candidate for dark
matter particles. We also demonstrated that development of perturbations in
baryons possesses an additional damping (with respect to the standard
Newtonian instability). However the complete analysis requires considering
the relativistic case which we present elsewhere.

As far as basic equations are concern, the generalization to the
relativistic (GR) case of our basic results seem to have not essential
difficulties (Boltzmann , fluid equations). The basic difficulty appears
when considering the generalization for the bias operator $B\left(
x,x^{\prime }\right) $ which reflects the polarizability of space. For
scalar wave equation in the geometric optics approximation such bias was
considered recently by \cite{KSS09}. However such an approximation works
only on sufficiently late stage of the evolution of the Universe (where the
Newtonian consideration works well), while for very early stage it is not
sufficient. This poses a rather serious problem for future research.

We point also out that for small astrophysical objects (e.g., jets, nuclei
of galaxies, etc.) one probably has to account for the dynamics of the
additional parameters of wormholes ($b_w,U$) which should essentially
complexify the theory presented.

The gas of wormholes represents an extremely reach (by physical effects) but
complex medium. Due to the existence of the polarizability (which has the
topological origin) it has all properties of a gravitational plasma. It
possesses also numerous additional non-trivial phenomena and we think it
worth saying that we have deal with a new state of matter.

\section{Acknowledgment}

This research was supported in some part ($e^{2}/\hslash c$) by RFBR
09-02-00237-a.

\end{document}